\documentclass[aps,pra,reprint,superscriptaddress]{revtex4-2}

\usepackage{soul}
\usepackage{graphicx}
\usepackage{amsmath}
\usepackage{amsfonts}
\usepackage{float}
\usepackage{xcolor}
\usepackage{xspace}
\usepackage[inline,shortlabels]{enumitem}
\usepackage{siunitx} 
\usepackage{xfrac} 
\usepackage[normalem]{ulem}
\usepackage[version=4]{mhchem} 

\begin{document}

\title{Effect of metal encapsulation on bulk superconducting\\ properties of niobium thin films used in qubits}

\author{Amlan Datta}
\affiliation{Ames National Laboratory, Ames, IA 50011, U.S.A.}
\affiliation{Department of Physics \& Astronomy, Iowa State University, Ames, IA 50011, U.S.A.}

\author{Kamal R. Joshi}
\affiliation{Ames National Laboratory, Ames, IA 50011, U.S.A.}

\author{Sunil Ghimire}
\affiliation{Ames National Laboratory, Ames, IA 50011, U.S.A.}
\affiliation{Department of Physics \& Astronomy, Iowa State University, Ames, IA 50011, U.S.A.}

\author{Bicky S. Moirangthem}
\affiliation{Ames National Laboratory, Ames, IA 50011, U.S.A.}
\affiliation{Department of Physics \& Astronomy, Iowa State University, Ames, IA 50011, U.S.A.}

\author{Makariy A. Tanatar}
\affiliation{Ames National Laboratory, Ames, IA 50011, U.S.A.}
\affiliation{Department of Physics \& Astronomy, Iowa State University, Ames, IA 50011, U.S.A.}

\author{Mustafa Bal}
\affiliation{Superconducting Quantum Materials \& Systems Division, Fermi National Accelerator Laboratory, Batavia, IL 60510, U.S.A.}

\author{Zuhawn Sung}
\affiliation{Superconducting Quantum Materials \& Systems Division, Fermi National Accelerator Laboratory, Batavia, IL 60510, U.S.A.}

\author{Sabrina Garattoni}
\affiliation{Superconducting Quantum Materials \& Systems Division, Fermi National Accelerator Laboratory, Batavia, IL 60510, U.S.A.}

\author{Francesco Crisa}
\affiliation{Superconducting Quantum Materials \& Systems Division, Fermi National Accelerator Laboratory, Batavia, IL 60510, U.S.A.}

\author{Akshay Murthy}
\affiliation{Superconducting Quantum Materials \& Systems Division, Fermi National Accelerator Laboratory, Batavia, IL 60510, U.S.A.}
\author{David A. Garcia-Wetten}
\affiliation{Department of Materials Science, Northwestern University, Evanston, Illinois 60208, U.S.A.}

\author{Dominic P. Goronzy}
\affiliation{Department of Materials Science, Northwestern University, Evanston, Illinois 60208, U.S.A.}
\author{Mark C. Hersam}
\affiliation{Department of Materials Science, Northwestern University, Evanston, Illinois 60208, U.S.A.}
\affiliation{Department of Chemistry, Northwestern University, Evanston, IL 60208, U.S.A.}

\affiliation{Department of Electrical \& Computer Engineering, Northwestern University, Evanston, IL 60208, U.S.A.}

\author{Michael J. Bedzyk}
\affiliation{Department of Materials Science, Northwestern University, Evanston, Illinois 60208, U.S.A.}

\affiliation{Department of Physics \& Astronomy, Northwestern University, Evanston, IL 60208, U.S.A.}

\author{Shaojiang Zhu}
\affiliation{Superconducting Quantum Materials \& Systems Division, Fermi National Accelerator Laboratory, Batavia, IL 60510, U.S.A.}

\author{David Olaya}
\affiliation{National Institute of Standards and Technology, Boulder, Colorado 80305, U.S.A.}

\author{Peter Hopkins}
\affiliation{National Institute of Standards and Technology, Boulder, Colorado 80305, U.S.A.}

\author{Matthew J. Kramer}
\affiliation{Ames National Laboratory, Ames, IA 50011, U.S.A.}

\author{Alexander Romanenko}
\affiliation{Superconducting Quantum Materials \& Systems Division, Fermi National Accelerator Laboratory, Batavia, IL 60510, U.S.A.}

\author{Anna Grassellino}
\affiliation{Superconducting Quantum Materials \& Systems Division, Fermi National Accelerator Laboratory, Batavia, IL 60510, U.S.A.}

\author{Ruslan Prozorov}
\email[Corresponding author: ]{prozorov@ameslab.gov}
\affiliation{Ames National Laboratory, Ames, IA 50011, U.S.A.}
\affiliation{Department of Physics \& Astronomy, Iowa State University, Ames, IA 50011, U.S.A.}

\date{7 February 2026}

\begin{abstract}
Niobium metal occupies nearly 100\% of the volume of a typical 2D transmon device. While the aluminum Josephson junction is of utmost importance, maintaining quantum coherence across the entire device means that pair-breaking in Nb leads, capacitive pads, and readout resonators can be a major source of decoherence. The established contributors are surface oxides and hydroxides, as well as absorbed hydrogen and oxygen. Metal encapsulation of freshly grown surfaces with non-oxidizing metals, preferably without breaking the vacuum, is a successful strategy to mitigate these issues. While the positive effects of encapsulation are undeniable, it is important to understand its impact on the macroscopic behavior of niobium films. We present a comprehensive study of the bulk superconducting properties of Nb thin films encapsulated with gold and palladium/gold, and compare them to those of bare Nb films. Magneto-optical imaging, magnetization, resistivity, and London and Campbell penetration depth measurements reveal significant differences in encapsulated samples. Both sputtered and epitaxial Au-capped films exhibit the highest residual resistivity ratio and superconducting transition temperature, as well as the lowest upper critical field, London penetration depth, and critical current. These results are in good agreement with the microscopic theory of anisotropic normal and superconducting states of Nb. We conclude that pair-breaking in the bulk of niobium films due to the disorder throughout the film, not just from the surface, is significant source of quantum decoherence in transmons. We also conclude that gold capping not only passivates the surface but also affects the properties of the entire film, significantly reducing the scattering rate due to defects likely induced by surface diffusion if the film is not protected immediately after fabrication.
\end{abstract}
\maketitle

\section{Introduction}

Niobium thin films are among the most viable candidates for superconducting quantum circuits \citep{Kjaergaard2020,Nielsen2010,Devoret2013,Grezes2016,Huang2020,Siddiqi2021,Alto22,Wenskat2022,Xiong2022,Ezratty2023}. Even though the essential part of superconducting qubits is typically an aluminum Josephson junction, niobium is used for the rest of the circuitry, including readout resonators, capacitor pads, and coupling lines. For a qubit operation, the entire device must be in a quantum coherent state. It is well established that surfaces and interfaces can be significant sources of decoherence \citep{Oliver2013,Wang2015,Dial2016,Gambetta2017,Pappas2011,Drimmer2025,Murthy2022,torres2024}. In particular, native niobium oxides, such as NbO, NbO$_{2}$, and Nb$_{2}$O$_{5}$, are believed to host two-level systems (TLS) and other pair-breakers, which contribute significantly to decoherence \citep{Leon2021,McDermott2009,Oliver2013,Wang2015,Dial2016,Read2023,goronzy2025}. Niobium can take on many oxidation states, resulting in non-stochiometric oxides. Many studies have been performed to either remove surface oxides or prevent their formation. This includes annealing in vacuum at high temperatures \citep{alex20}, wet etching with HF \citep{Alto22,kopas2024}, and forming a niobium nitride layer with nitrogen plasma passivation at the metal-air interface \citep{Zheng22}. Currently, there is a growing interest in the metal encapsulation of freshly deposited Nb surfaces, ideally in-situ without breaking the vacuum in the deposition chamber, so that the sample is never exposed to oxygen. Metal over-layers several nanometers thick, such as ruthenium, gold, palladium/gold, aluminum, and tantalum, have been shown to improve the performance of superconducting devices \citep{Senthil24,Chang2024,Ory2024,Bal2024}. Specifically, the coherence of Ta-capped Nb films was improved by almost 5 times (increased from 40\,$\mu$s to 200\,$\mu$s) \citep{Bal2024}. However, the microscopic origin of this improvement is not yet understood. Many studies focus on surfaces, finding significant differences between capped and bare films \cite{Bal2024, Chang2024, Senthil24}, and often conclude that the metallic over-layers significantly affect the surface but not the interior of the film. Here, we demonstrate that in addition to the alteration of the surface, the bulk properties are also significantly affected by capping layers. We note that while our films are ``thin", they are far from the ultra-thin limit, which in niobium is known to less than approximately 25\,nm, when all ``bulk" properties start changing rapidly: $T_c$ and upper critical field decrease, while resistivity, and London penetration depth increase \citep{ilin2004,pinto2018}. 

In this paper, we investigate the impact of surface encapsulation on the bulk superconducting properties of 110-120\,nm thick Nb films grown epitaxially (control (reference) uncapped films and Au-capped films) and 150-155\,nm thick sputtered (control (reference) uncapped films, Au-, and PdAu-capped film) onto a sapphire substrate, as described in section \ref{section:samples}. The multi-modal characterization of the same films using magneto-optical imaging, magnetization, resistivity, and London and Campbell penetration depths shows that encapsulation not only affects the surface but also significantly changes the bulk response of capped films. 

\section{Results}

\subsection{Resistivity}

\begin{figure}[!tbh]
\includegraphics[width=0.95\linewidth]{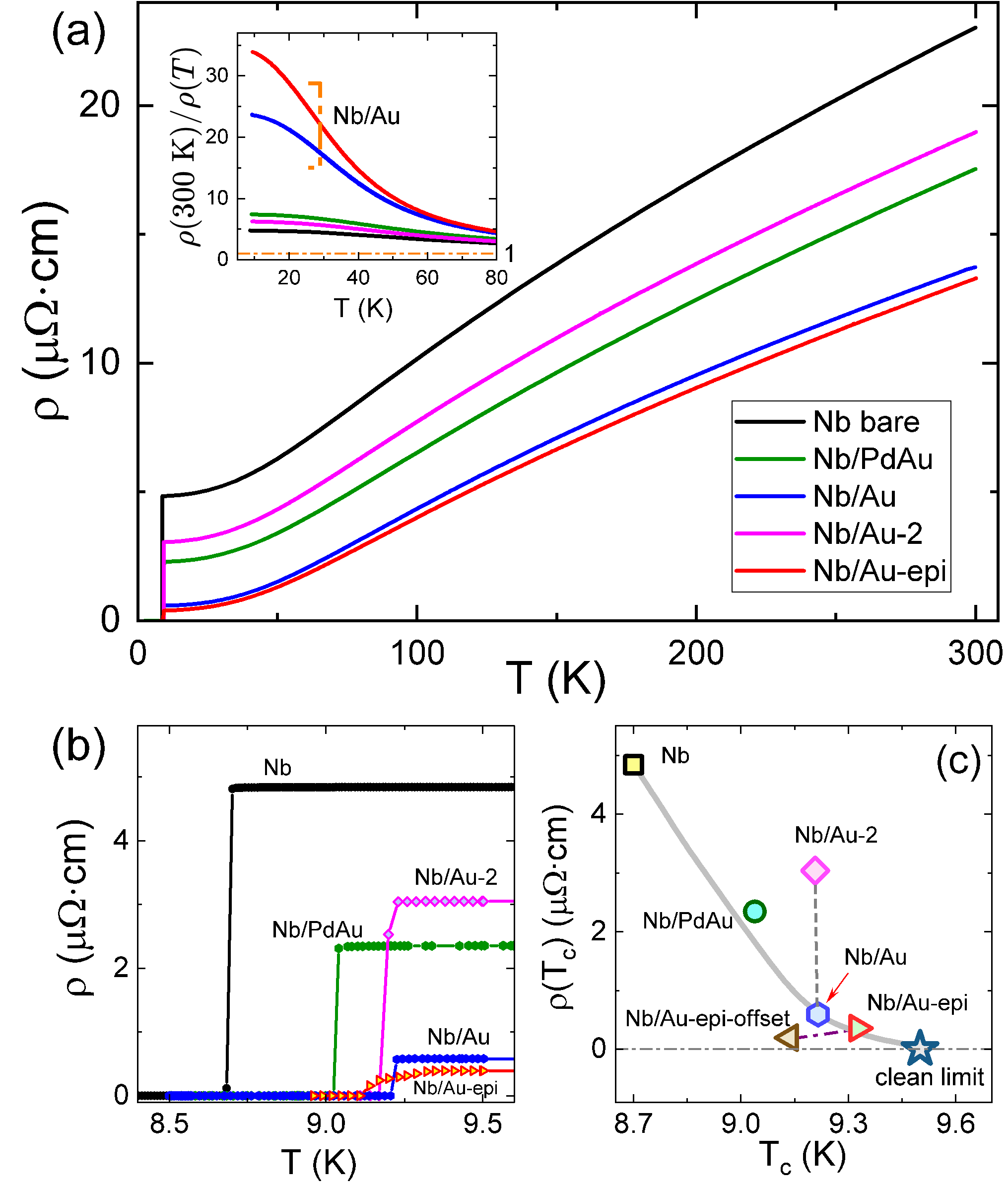} 
\caption{\textbf{Electrical resistivity.} (a) Full temperature scale resistivity of all studied films. Inset in (a): residual resistivity ratio, $RRR(T)=\rho(300\;\text{K})/\rho(T)$ showing distinctly higher values in Au-capped films. (b) Zoom in on the superconducting transition, $T_c$. Au-capped films show the highest $T_{c}$. (c) Resistivity at $T_{c}$ as a function of $T_{c}$ showing monotonic behavior except for the un-optimized Au-capped film, Nb/Au-2.}
\label{fig:resistivity} 
\end{figure}

Figure~\ref{fig:resistivity}(a) shows the temperature-dependent resistivity, $\rho (T)$, in the full temperature range i.e. from 300\,K down to 7\,K. The inset in Fig.\ref{fig:resistivity}(a) shows the ratio, $\rho (300\,\text{K})/\rho (T)$. At low temperatures, there are two primary contributions to resistivity: electron-electron scattering ($\rho \sim T^2$ in the Fermi liquid regime) and scattering off defects and impurities, which is temperature independent. Figure~\ref{fig:resistivity}(b) zooms into the region near $T_c$, showing temperature-independent $\rho (T)$. Therefore, the values of $\rho (T_c)$ serve as a good proxy for the residual resistivity, $\rho (0)$; hence, the lower limit of $\rho (300\,\text{K})/\rho (T_c)$ provides a good approximation for the residual resistivity ratio, $RRR \equiv \rho(300\;\text{K})/\rho(0)$, which is commonly used to gauge the scattering rate due to defects and impurities. For comparison, the cleanest niobium samples (bulk) achieve $RRR \approx 90000$ \citep{Koethe2000}, but the typical values in thin films are much lower, ranging from 5 to 50 \citep{Joshi2023,oh2024}. Among the studied films, Au-capped films, including Nb/Au and Nb/Au-epi, have the highest $RRR$ values, indicating significantly lower disorder than other films. We note that the highest onset $T_c$ and the highest $RRR$ is observed in a Nb/Au-epi film. However, this film also shows a more smeared (e.g., gradual) transition curve. The lowest $\rho (T_c)$ suggests that the smearing does not arise from an elevated density of scattering centers and is likely due to compositional or structural inhomogeneity. As we report below, this film also exhibits a larger value of the London penetration depth, $\lambda (0)$, which is consistent with this hypothesis. On the other hand, the offset temperature, when $\rho(T)=0$, remains within the range of other Au-capped films and is significantly higher than that of the rest of the samples. 

The studied films have thicknesses ranging from 110\,nm to 163\,nm, and detailed information is provided later in the paper in section \ref{section:samples}. For reference, the resistivity of Nb at 300 K is \qty{15}{\micro\ohm\cm} and \qty{5e-4}{\micro\ohm\cm} at 4.5 K \citep{Koethe2000}. Therefore, the resistivity of the capping layer, which ought to be added in parallel with the niobium film, contributes negligibly, given their thickness of less than 10 nm.

 Figure~\ref{fig:resistivity}(c) shows $\rho(T_{c})$ vs. $T_{c}$. In our case, the significant variation of the transition temperature must be predominantly due to differences in scattering center density, not the grain size \citep{Tanatar2022, Joshi2023}. If the granular structure were important, it would result in a significant smearing of the transition curves, $\rho(T)$, but we observe sharp transitions in all film types. According to Ref.\citep{Zarea2023}, Nb is a multiband superconductor and has an anisotropic gap structure; therefore, it is suppressed by nonmagnetic disorder, which has been verified experimentally by a proton irradiation study \citep{Tanatar2022}. The horizontal dashed line shows a theoretical clean limit, \qty{5e-4}{\micro\ohm\cm}  \citep{Koethe2000}, which visually appears as zero in the figure. The blue star in Fig.~\ref{fig:resistivity}(c) shows the corresponding $T_c$ in the clean limit \citep{Prozorov2022}. The solid gray line represents the suppression of the transition temperature with disorder, as estimated from theory \citep{Zarea2023}.

\subsection{Quasiparticle spectroscopy - superfluid density}

\begin{figure}[!tbh]
\includegraphics[width=0.95\linewidth]{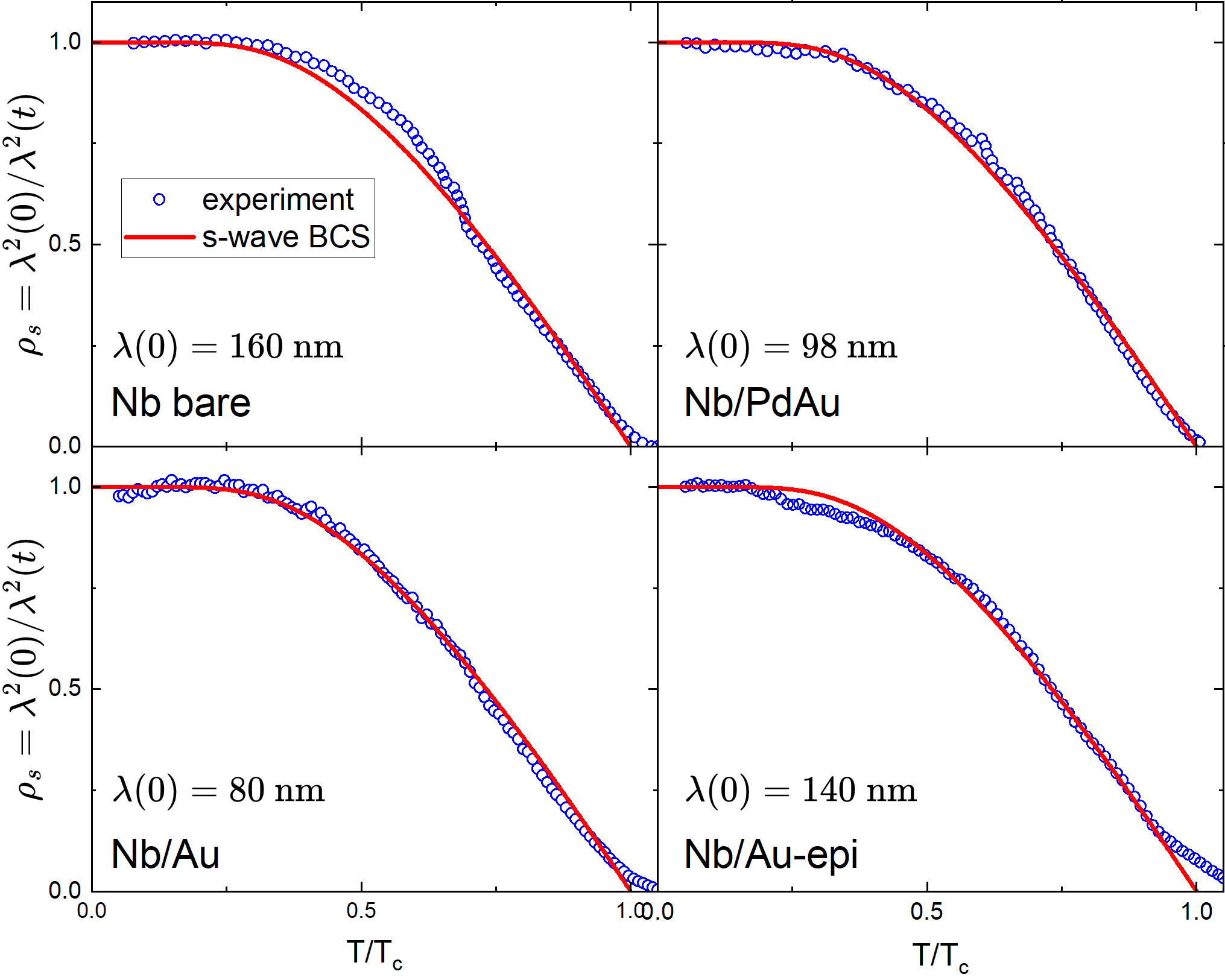} 
\caption{\textbf{Superfluid density.} Normalized superfluid density, $\rho_{s}=(\lambda(0)/\lambda(T))^{2}$ (symbols), calculated from the measured London penetration depth, $\lambda(T)$. The solid red curves show the isotropic weak-coupling BCS theory. The only free parameter to adjust the experimental data to match the theory was $\lambda(0)$, and this is how it was obtained for Fig.\ref{fig:Hc2_Tc}.}
\label{fig:superfulid_density} 
\end{figure}

Achieving long coherence times and high quality factors in superconducting quantum circuits is hindered by quasiparticles from thermal and other pair-breaking mechanisms \citep{Lutchyn05,Siddiqi2021}.  It is important to know the presence of quasiparticles in the superconducting regime to ensure the mitigation of decoherence. The density of the quasiparticles is proportional to $\sqrt{\pi\Delta (0)/2k_B T}\exp(-\Delta (0)/k_B T)$, which is proportional to the variation of the London penetration depth, $\Delta\lambda(T)$. The high precision measurement of $\Delta\lambda(T)$ is achieved using the tunnel diode resonator (TDR) technique, as described in section \ref{section:TDR}, is also used to calculate the superfluid density, $\rho(T)=1/(1+\Delta\lambda(T)/\lambda(0))^{2}$, which can be compared to the expectations from theory \citep{Prozorov2006,Prozorov2011,Prozorov2021}. The zero-temperature value of the penetration depth, $\lambda(T=0)$, was extracted by adjusting it as the only free parameter to fit to the full temperature range of the weak-coupling theoretical BCS (Bardeen–Cooper–Schrieffer) curve \citep{Bardeen1957}, shown in Fig.~\ref{fig:superfulid_density}. Note that if the temperature dependence of the measured penetration depth were inherently different from the BCS curve, a mere adjustment of $\lambda(T=0)$ would not make it fit. We therefore conclude that, within the experimental uncertainty, all studied films follow the standard BCS curve. Theoretically, Nb is a multiband superconductor with somewhat stronger coupling and an anisotropic order parameter \citep{Zarea2023}. However, these deviations from isotropic BCS have little effect on the full-range temperature dependence of the superfluid density, which is determined by the integral of the density of states over the entire Fermi surface and all quasiparticle energies. Direct fitting to a theoretical stronger coupling curve shows that the extracted $\lambda(0)$ remains practically unchanged since the entire curve is used \citep{ghimire2024i}.

\subsection{Magnetization}

\begin{figure}[!tbh]
\includegraphics[width=0.95\linewidth]{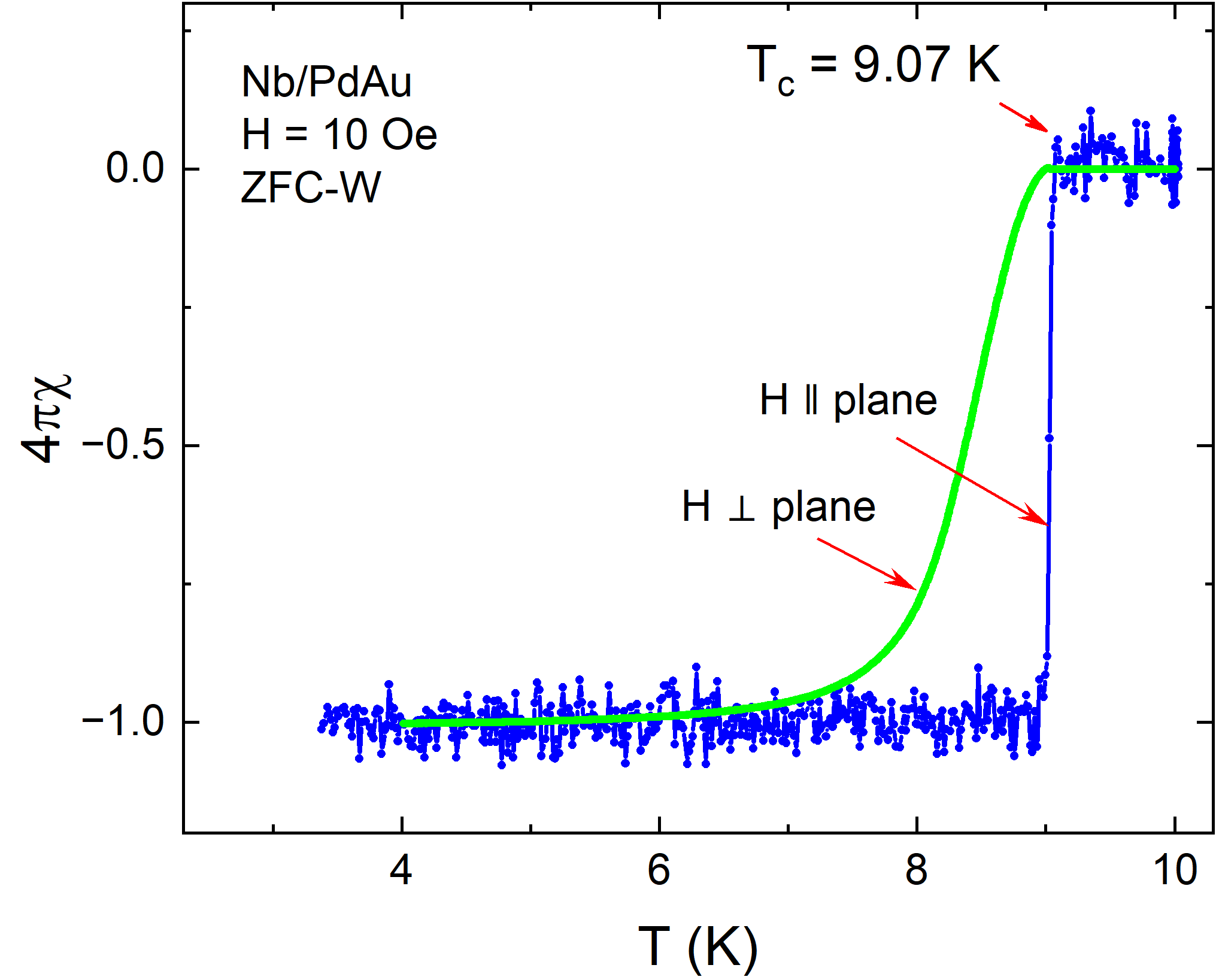} 
\caption{\textbf{Parallel vs. perpendicular orientation.} Temperature dependence of magnetic susceptibility, $\chi(T)$, of PdAu-capped Nb film. The green symbols correspond to a magnetic field applied perpendicular to the film’s plane. Blue symbols show the case when a magnetic field was applied parallel to the film’s plane. Note a significant difference in noise level due to extreme demagnetization.}
\label{fig:chi} 
\end{figure}

A typical parameter used to assess the quality of films in the superconducting regime is its  magnetization, $M$. The interaction between Abrikosov vortices and material defects (vortex pinning) plays a crucial role in determining the magnetic behavior of type-II superconductors and the maximum dissipation-free supercurrent they can support. The irreversibility observed in magnetic hysteresis loops, $M(H)$, is a measure of its pinning strength. However, there is a significant concern with high demagnetizing effects when the orientation of the magnetic field is perpendicular to the film's plane \citep{Prozorov2018}. In this case, it is very difficult to infer the intrinsic pinning properties and to compare different films. This is demonstrated in Fig.~\ref{fig:chi}, where DC magnetic susceptibility (measured by QD PPMS VSM)\ref{fn:PPMS} is shown for the PdAu-capped Nb film as a function of temperature in two orientations.

A magnetic field was applied in both perpendicular and parallel directions to the same film in separate runs. For comparison, the curves were normalized to a range of -1 to 0. The green curve in Fig.~\ref{fig:chi} is for a magnetic field applied perpendicular to the film surface, while the blue dots show the data for a magnetic field applied in the parallel direction. Clearly, the commonly used constant geometric factor cannot be applied to correct for such a substantial demagnetizing in this extremely thin film geometry, in part because the effect depends on temperature, $\chi=\chi_0/(1+N\chi_0)$, where $\chi_0$ is the magnetic susceptibility of an infinite sample and $N$ is the demagnetizing factor \citep{Prozorov2018}. 

\begin{figure}[!tbh]
\includegraphics[width=0.95\linewidth]{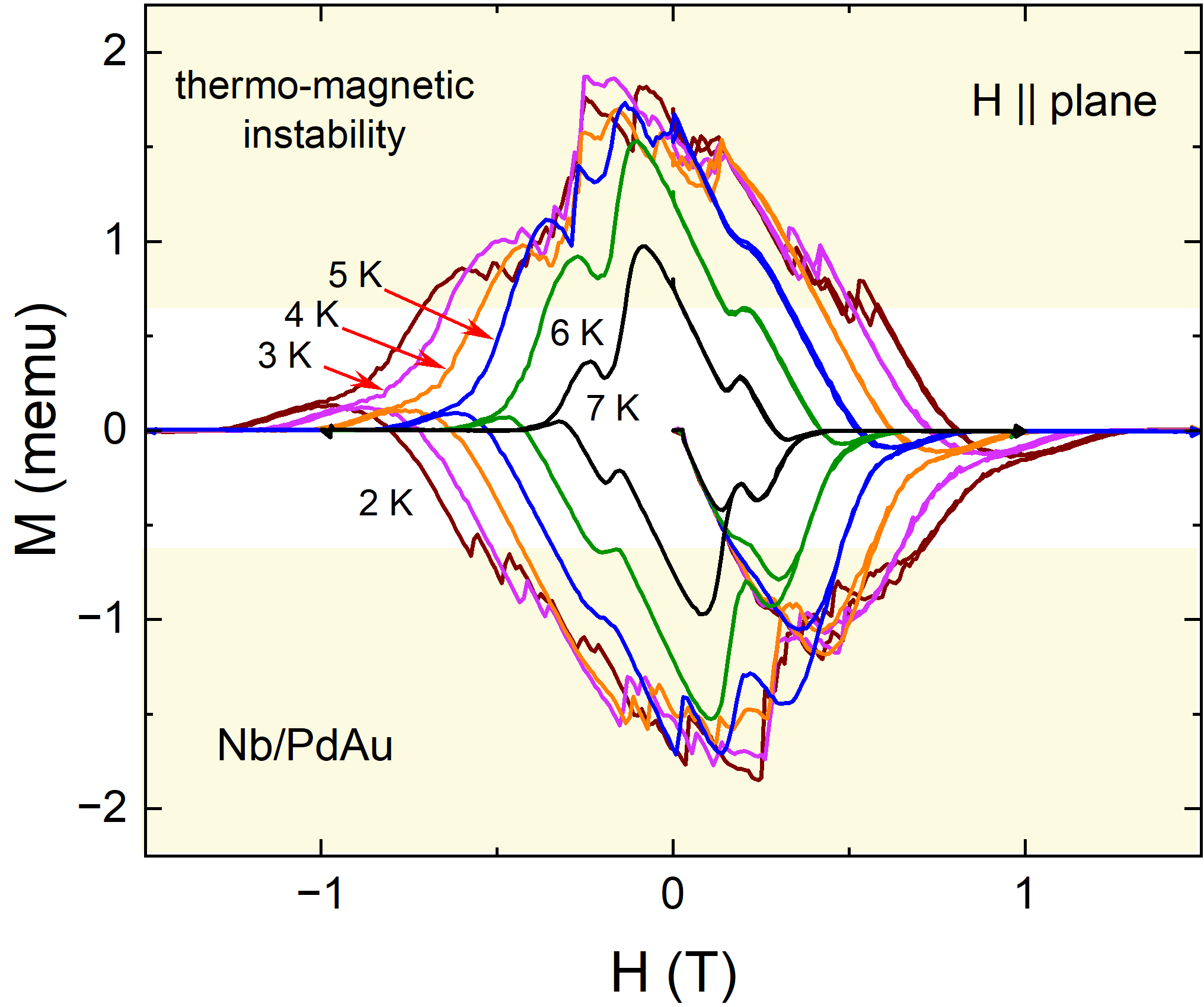} 
\caption{\textbf{Thermo-magnetic flux avalanches.} Magnetization loops (raw data in memu) measured at indicated temperatures in a parallel orientation in PdAu-capped Nb film. The vertical jumps appearing above a certain threshold (shown by light shaded areas) are well-known in niobium films, occurring due to thermomagnetic instabilities (a.k.a. vortex avalanches) \citep{Duran1995, Welling2004}.}
\label{fig:magnetization_PdAu} 
\end{figure}

Figure~\ref{fig:magnetization_PdAu} shows the measured magnetic moment (raw data in memu) as a function of the magnetic field applied parallel to the film surface in the Nb/PdAu film. The field sweeps are conducted at various temperatures, ranging from 2 K to 7\,K in 1 K increments. At lower temperatures, we observe pronounced jumps in magnetization, which cease at higher temperatures. This is a well-known behavior for Nb films: thermomagnetic instabilities, also known as vortex avalanches \citep{BlancoAlvarez2019, Duran1995, Welling2004}.

\begin{figure}[!tbh]
\includegraphics[width=0.95\linewidth]{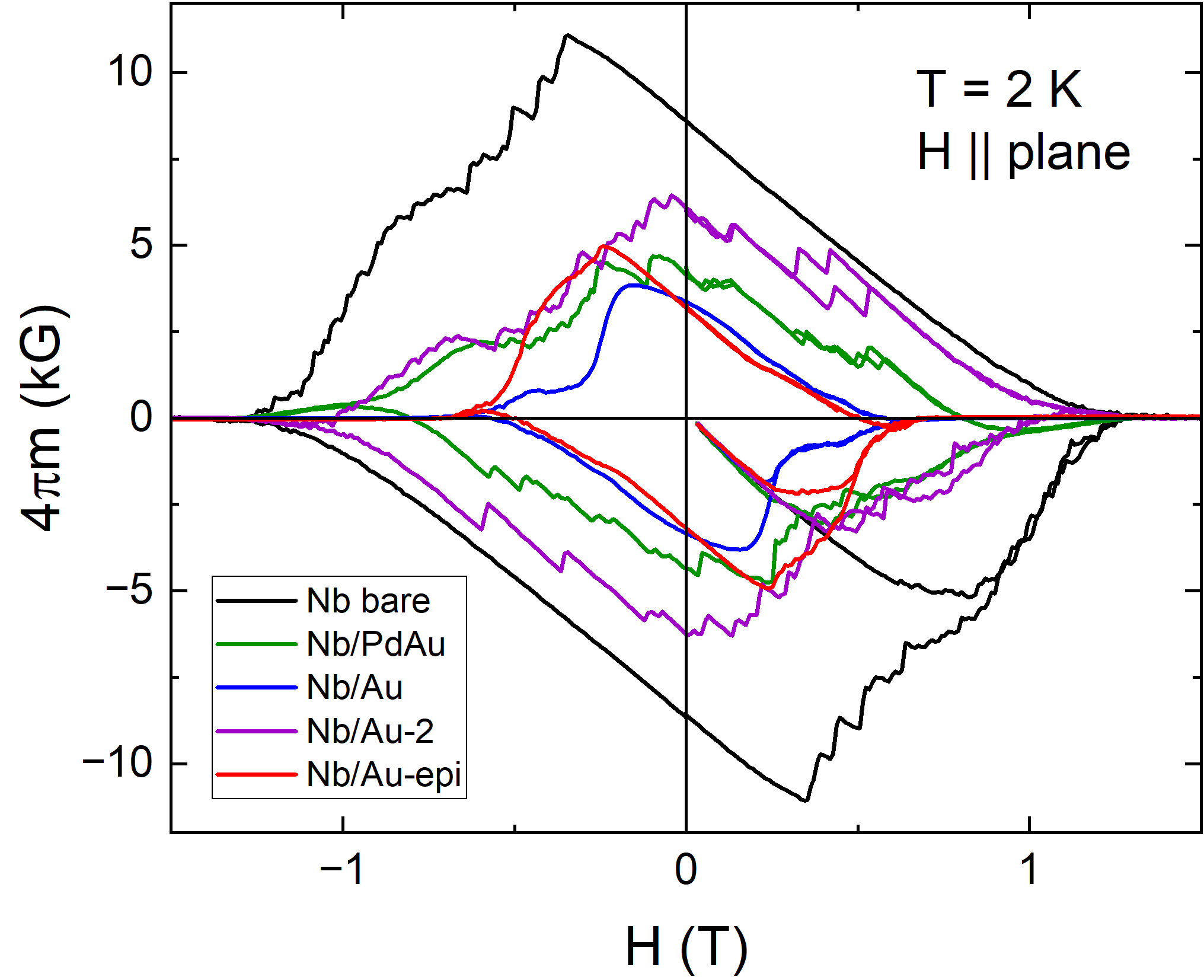} 
\caption{\textbf{DC magnetization.} Comparison of normalized magnetic hysteresis loops, $4\pi m(H)$ (in gauss), at 2 K in the studied films. The magnetic field was applied parallel to the sample plane. The hysteresis is the largest in a bare Nb film and the lowest in an Au-coated film, implying stronger pinning in the former and weaker pinning in the latter.}
\label{fig:magnetization_all} 
\end{figure}

To compare different films, the measured magnetic moment is normalized so that the initial slope (after cooling in zero field) is $dm/dH=-1$. Figure~\ref{fig:magnetization_all} shows a comparison of the normalized magnetization (in gauss) recorded at 2 K for all the films under study. We observe similar jumps in all the films, except for the Au-capped ones, whether they are sputtered or epitaxially grown. These abrupt magnetic flux jumps occur when the critical current density is sufficiently high, and the heat generated by rapidly moving vortices has no effective means of dissipating, leading to a catastrophic thermomagnetic runaway. In some cases, this can even cause the complete collapse of the critical state, resulting in measurable jumps in the sample's temperature \citep{Prozorov2006a}. Of all technologically important materials, niobium is highly susceptible to these thermomagnetic instabilities \citep{Joshi2023,Datta2024}. Differences in pinning strength (hence, critical current density) and metal-substrate thermal coupling result in distinctly different hysteresis loops.

\subsection{Upper critical field, $H_{c2}(0)$, and London penetration depth, $\lambda(0)$}
Two other experimental parameters sensitive to disorder are the upper critical field, $H_{c2}(0)$, and the  London penetration depth, $\lambda(0)$, both in the $T \to 0$ limit. The disorder is parameterized by the dimensionless scattering rate, $\Gamma=\hbar/(2 \pi \tau k_B T_{c0})$, where $\tau$ is the scattering time and $T_{c0}$ is the clean-limit transition temperature \citep{AbrikosovGorkov1960ZETF}. There are several approaches to estimate $\Gamma$. Here, we note that the initial suppression of the actual transition temperature is given by $T_{c} \left( \Gamma \right)=T_{c0}(1-A\tfrac{\pi^2}{4}\Gamma)$, where $A\approx 0.037$ is the anisotropy parameter for Nb \citep{Zarea2023,Tanatar2022}. Therefore, $\Gamma =\tfrac{4}{A\pi^2T_{c0}}\left( T_{c0}-T_c \right)$, and in practice, it is convenient to plot different $\Gamma$-dependent parameters vs. the actual $T_c(\Gamma)$, which decreases linearly with $\Gamma$.

\begin{figure}[!tbh]
\includegraphics[width=0.95\linewidth]{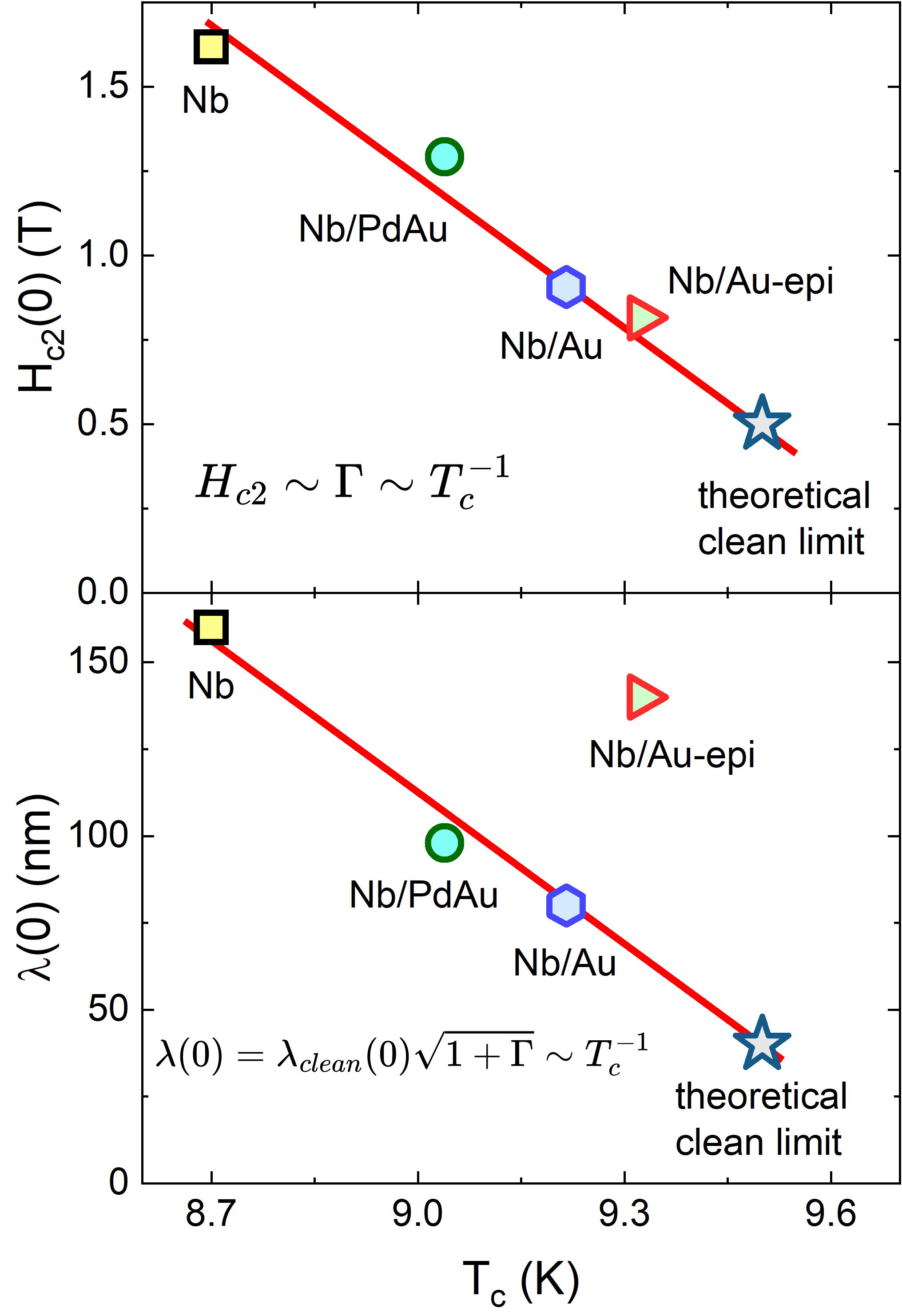} 
\caption{\textbf{The upper critical field and London penetration depth.} The upper critical field, $H_{c2}(T=0)$, (top panel) and London penetration depth $\lambda(T=0)$ (bottom panel), plotted versus superconducting transition temperature, $T_{c}$. The $H^0_{c2}$ was obtained from the Helfand-Werthamer fit of $H_{c2}(T)$ and $\lambda(0)$ was estimated as part of the analysis for Fig.~\ref{fig:superfulid_density}. See text for details. Both quantities are expected to \textit{increase} linearly with the \textit{decrease} of  $T_{c}$, and the same is observed in our experiment.} 
\label{fig:Hc2_Tc} 
\end{figure}

The $\Gamma$-dependent upper critical field is $H_{c2}\left( \Gamma \right)=H^0_{c2}(1+3\Gamma/4)$, where $H^0_{c2}$ is the clean-limit value \citep{Zarea2023,Tanatar2022}, which has been obtained from the Helfand-Werthamer fit of $H_{c2}(T)$ data \citep{HW1966,Prozorov2024}. The top panel of Fig.~\ref{fig:Hc2_Tc} shows $H_{c2}(0)$ vs. $T_{c}$. As expected, the most disordered bare Nb film has the highest upper critical field, while the Au-capped epitaxial film, Nb/Au-epi, has the lowest $H_{c2}$ among the studied samples. For completeness, we added the theoretical clean limit value for niobium, $H^0_{c2}=5$~kOe \citep{Finnemore1966}. The bottom panel of Fig.~\ref{fig:Hc2_Tc} shows the magnitude of the London penetration depth, $\lambda(0)$, obtained from fitting the data to an $s$-wave BCS curve, as explained in the discussion of Fig.\ref{fig:superfulid_density}. The $\lambda(0)$ as a function of transition temperature is consistent with the theory \citep{Zarea2023}. 

\subsection{Magneto-optical visualization of the magnetic flux}

\begin{figure*}[tbh]
\includegraphics[width=0.95\linewidth]{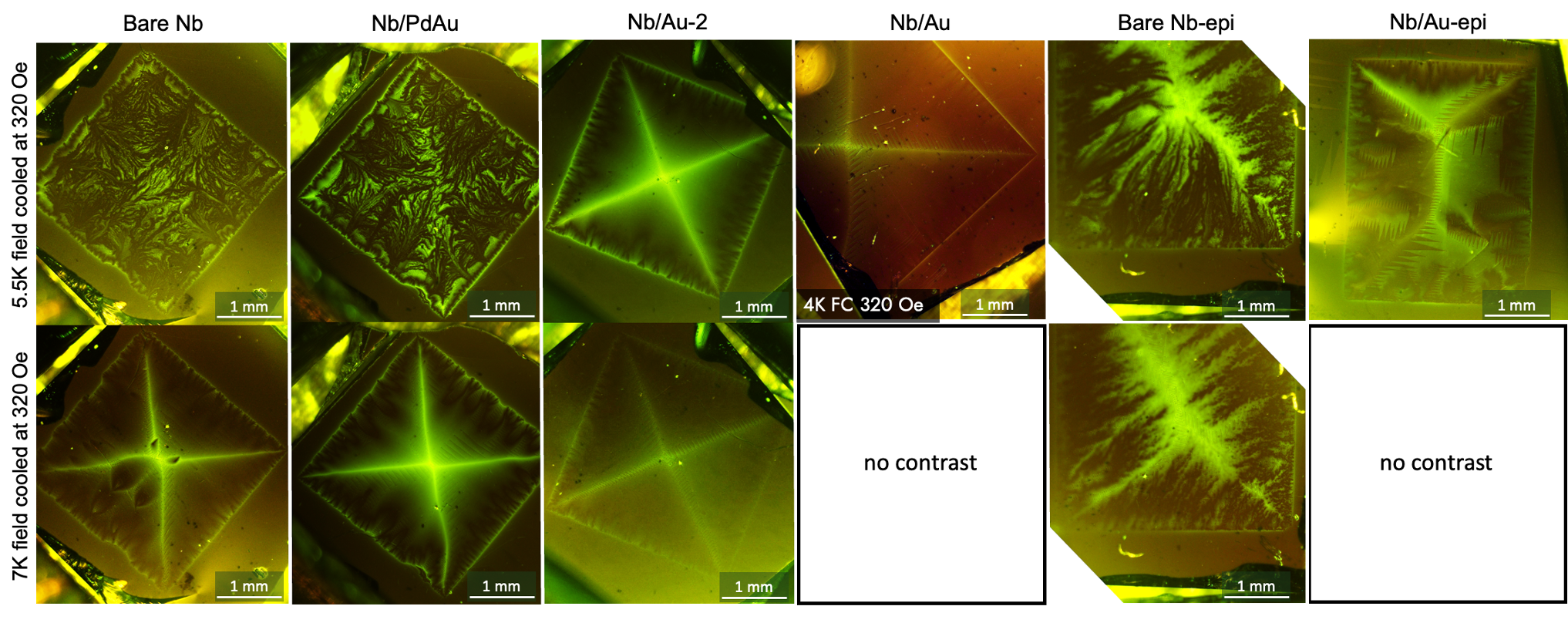} 
\caption{\textbf{Magneto-optical visualization: remanent state after FC.} Magneto-optical Faraday images of field cooled (FC) Nb thin films in the remanent state, magnetic field was removed at 5.5\,K (upper panel) and, in a separate run, at 7\,K (bottom panel). At 5\,K temperatures, apart from Au-capped films, the films show signatures of thermo-magnetic dendritic avalanches trapped inside the sample. At 7\,K, the trapped magnetic flux is uniform in a rooftop-like shape, as described by the Bean model \citep{Bean1962, Brandt1995}.}
\label{fig:MO_FC} 
\end{figure*}

In this section, we discuss the magneto-optical imaging of niobium films. We explored the processes of magnetic flux penetration at low temperatures after cooling in zero-field (ZFC) and flux trapping after cooling in a magnetic field from above the superconducting transition and turning the field off. In all of the MO images, brighter contrast corresponds to higher $B_{z}$, and black regions correspond to no magnetic induction in the superconducting state.
\begin{figure*}[tbh]
\includegraphics[width=0.95\linewidth]{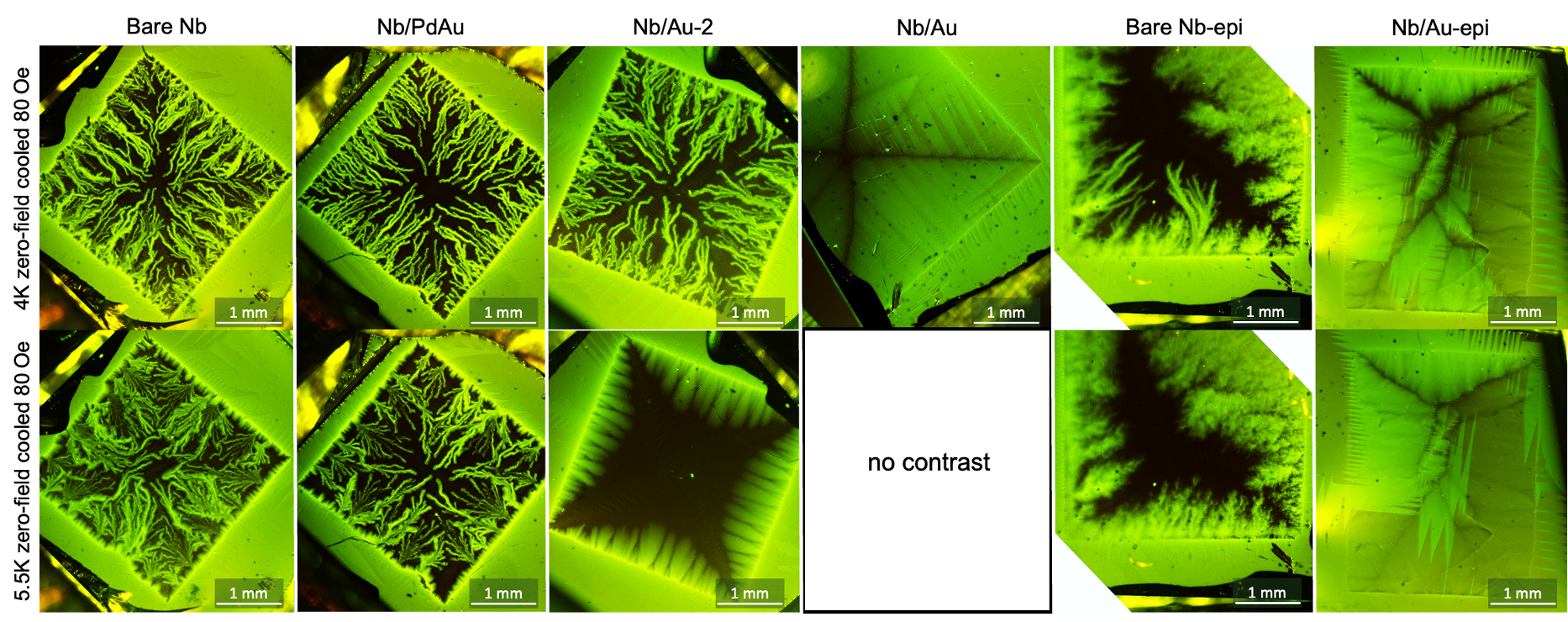} 
\caption{\textbf{Magneto-optical visualization: ZFC.} Faraday-effect based magneto-optical images of Nb films after zero-field cooling to 4\,K (upper panel) and, in another run, to 5.5\,K (lower panel) after which a magnetic field of 80\,Oe was applied. The first four columns show sputtered films, and the other two show epitaxially grown films. Except for the Au-capped thin film, dendritic flux avalanches are observed in all films. Judging by the flux penetration depth, the images show that Nb/Au and Nb/Au-epi films have the lowest flux pinning, consistent with being cleaner as strongly suggested by Fig.~\ref{fig:Hc2_Tc}. No contrast implies no response from the film, as the superconducting state has already collapsed.}
\label{fig:MO_ZFC_lowT} 
\end{figure*}
The upper row of Fig.~\ref{fig:MO_ZFC_lowT} describes the behavior of the Nb films after cooling in zero field to 4\,K and, in another run, to 5.5\,K, after which a magnetic field of 80\,Oe was applied. A significant observation is that the bare film, Nb/PdAu film, and one of the sputtered Au-capped films clearly exhibit thermo-magnetic avalanches when a field is applied. On the other hand, as shown in the two rightmost columns of Fig.~\ref{fig:MO_ZFC_lowT}, the epitaxially grown bare Nb thin film, although it exhibits dendritic avalanches at low temperatures, it does not display this behavior when capped with Au. A very smooth and homogeneous flux penetration is observed instead. It is worth noting that the flux pinning landscape differs somewhat in the case of an epitaxially grown film as compared to sputtered films. In contrast to sputtered films, MBE films are not granular, which apparently suppresses the thermo-magnetic avalanches.
To reveal the intrinsic pinning landscape, we need to arrest these avalanches. They appear when the critical current density exceeds a certain threshold value. Therefore, we need to try to raise the temperature sufficiently. Figure~\ref{fig:MO_ZFC_highT} shows magneto-optical images at 6\,K (top row) and 7\,K (bottom row), where we were able to eliminate the dendrites and observe the ``true" flux pinning landscape. At 6\,K, the epitaxial Au-capped films exhibit more flux penetration into the center of the film as compared to other films at the same field and temperature. At 7\,K, the signal from this film became unresolvable. Overall, in Au-capped films, whether sputtered or epitaxially grown, the pinning is the lowest.
\begin{figure*}[!tbh]
\includegraphics[width=0.95\linewidth]{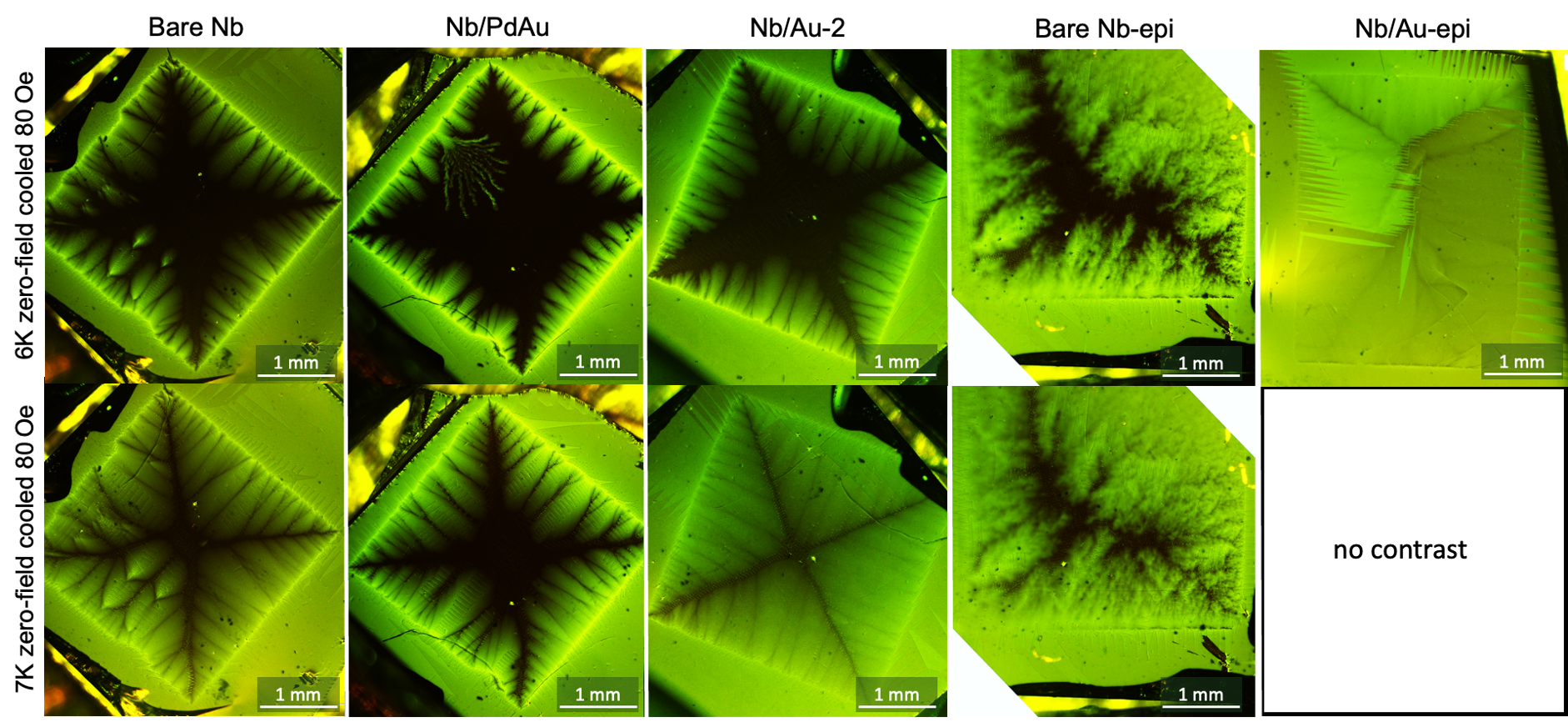} 
\caption{\textbf{Magneto-optical visualization: ZFC.} Comparison of the magnetic flux penetration in the studied films, similar to Fig.~\ref{fig:MO_ZFC_lowT}, but at higher temperatures, 6\,K (upper row) and 7\,K (bottom row), showing no thermo-magnetic instabilities. In the epitaxially grown Au-capped film, at 7\,K, the irreversible Bean critical state collapsed already in an applied magnetic field of 80\,Oe, resulting in no contrast to show.}
\label{fig:MO_ZFC_highT} 
\end{figure*}
To complete the imaging, Fig.~\ref{fig:MO_FC} shows the results of a field-cooled experiment in which a sample is cooled in a certain magnetic field to a target temperature; then, the applied field is removed, and magneto-optical images are taken. In Fig.~\ref{fig:MO_FC}, the applied magnetic field was 320\,Oe, and the films were cooled to  5.5\,K and, separately, to 7\,K. Similarly to Fig.\ref{fig:MO_ZFC_lowT}, the dendritic structure is present at 5.5\,K in bare niobium and PdAu-capped films, whereas Au-capped films do not show such signatures; at higher temperatures, we observe a clean Bean-model structure of trapped flux \citep{Bean1962}. However, in Au-capped films, the signal decreases significantly. Therefore, instead of 5.5\,K, we show only the lower temperature of 4\,K, at which the flux structure could be resolved.
\subsection{Campbell penetration depth and true critical current}
\begin{figure}[!tbh]
\includegraphics[width=0.95\linewidth]{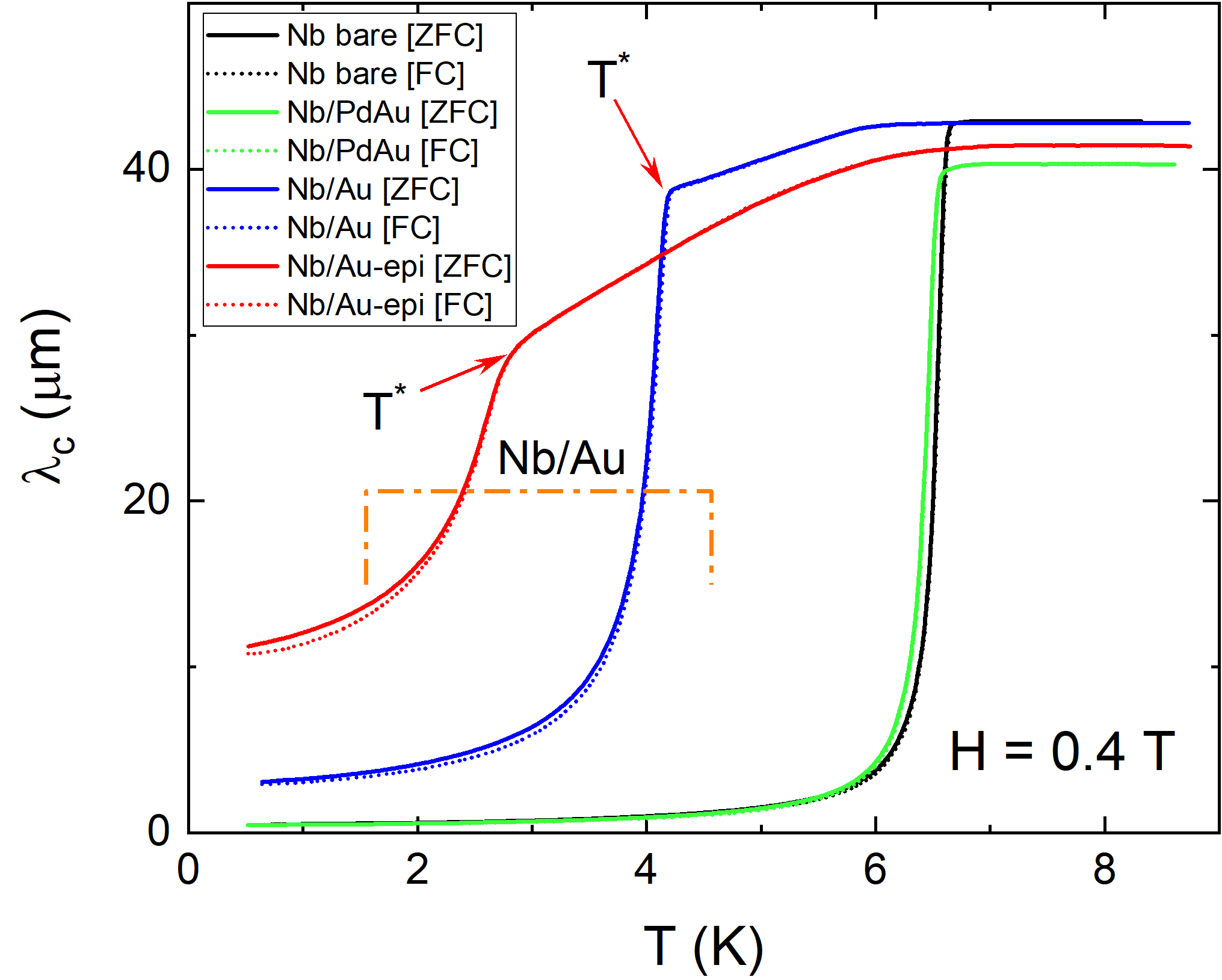} 
\caption{\textbf{Campbell penetration depth, $\lambda_C(T)$.} Measurements of $\lambda_C(T)$ at $H=0.4\;\text{T}$. Solid lines show ZFC measurements on warming, dotted lines show FC measurements on cooling. In addition to the expected lower $H_{c2}$, Au-capped films reveal a distinct kink at lower temperatures, likely due to the proximity effect with the non-superconducting gold layer. The larger the magnitude of the Campbell depth is, the lower the critical current, consistent with magneto-optical and magnetization measurements.}
\label{fig:campbell} 
\end{figure}

Magnetization and magneto-optical images provide information about ``persistent" current, which is a combination of the true critical current and magnetic relaxation \cite{sun2015}. The true critical current provides direct access to the pinning landscape; hence, it reflects the disorder in the studied films. We use measurements of Campbell penetration depth to extract the true critical current. When a sufficiently small amplitude AC magnetic field is applied to a sample with pinned vortices, such that they oscillate inside their potential wells, the AC perturbation is exponentially attenuated over a characteristic distance known as the Campbell penetration length, $\lambda_{C}$ \citep{Campbell1969,Kim2021,Ghimire24}. This parameter is conceptually similar to the London penetration depth, $\lambda$, but applies in the mixed state where vortices are present due to the DC magnetic field. The overall linear AC response is given by the measured total penetration depth,  $\lambda_{m}^{2}=\lambda^{2}+\lambda_{C}^{2}$, combining the contributions from both Meissner screening (via $\lambda$) and vortex lattice damping (via $\lambda_{C}$) \citep{Brandt1995}. 
Figure~\ref{fig:campbell} shows the Campbell penetration depth for the films. The comparison is provided for an externally applied field of $H=0.4\;\text{T}$ in TDR. Solid lines represent zero-field-cooled (ZFC) responses, measured upon warming, while dotted lines represent field-cooled (FC) measurements, with a 0.4~T external field applied in both cases. There is a significant difference between bare Nb and Au-coated films. The $T_c(H)$ is clearly suppressed in these films because the $H_{c2}(T)$ is lower, consistent with Fig.~\ref{fig:Hc2_Tc}. Furthermore, in the Au-capped films, we observe the appearance of a distinct kink at the temperature marked $T^*$. This is most likely due to the proximity effect with the gold capping layer, indicating good cohesion with niobium metal \citep{Prozorov_Proximity}. Quantitatively, a larger Campbell penetration depth implies a lower critical current density, $j_c=r_p H_0/\lambda^2_C$, where $r_p \approx \xi$, the coherence length. This is consistent with our analysis of magneto-optical images shown in Fig.\ref{fig:MO_ZFC_lowT} and \ref{fig:MO_ZFC_highT}.

\begin{figure}[!tbh]
\includegraphics[width=0.95\linewidth]{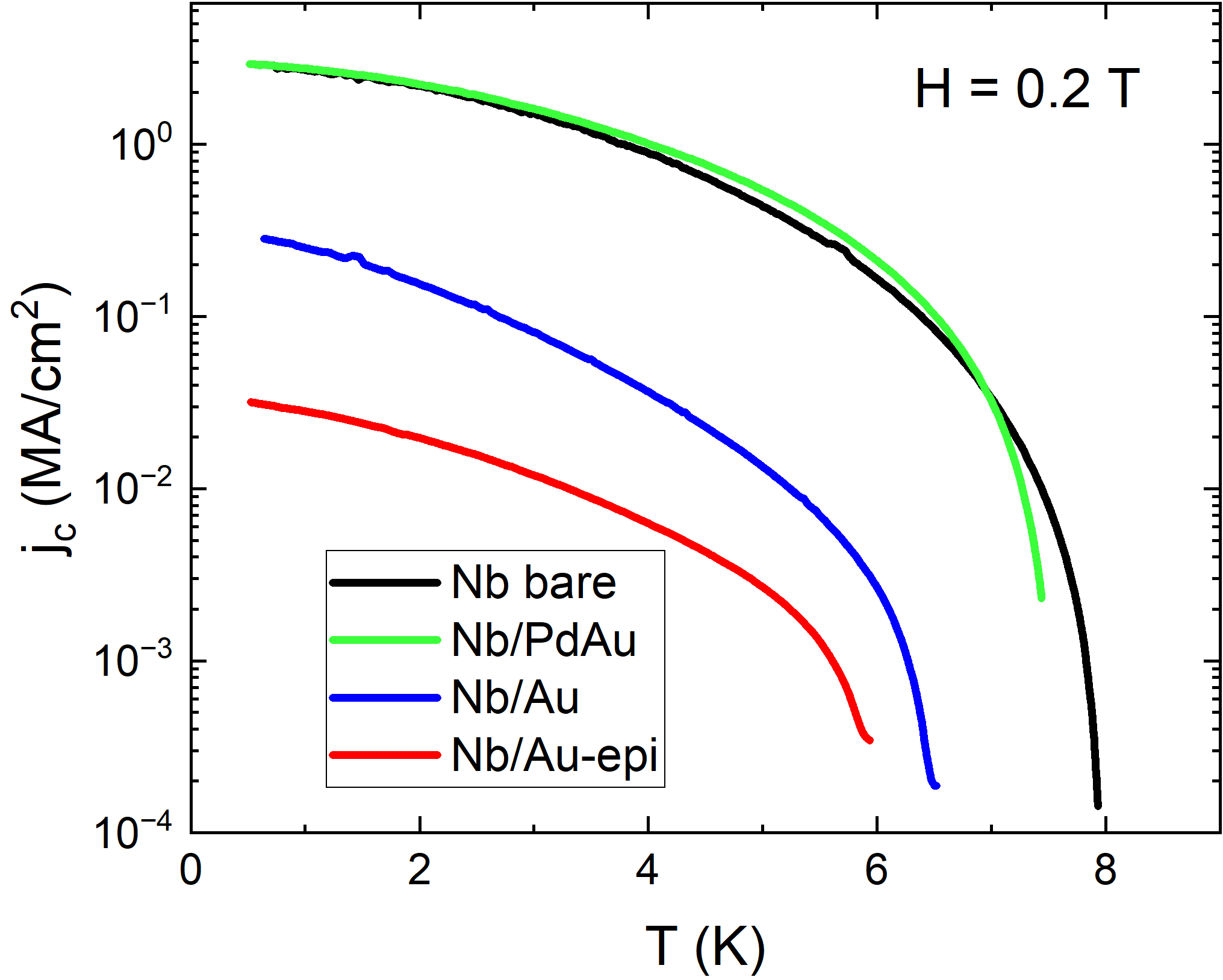} 
\caption{\textbf{Unrelaxed critical current density, $j_c(T)$.}  Temperature -dependent $j_c(T)$, evaluated at $H=0.2\;\text{T}$. Note a logarithmic vertical scale. }
\label{fig:jc_T} 
\end{figure}

The picture becomes even clearer when we examine the temperature-dependent critical current density, $j_{c}(T)$, extracted from the Campbell length. Figure~\ref{fig:jc_T} shows $j_{c}(T)$ for all films at an external magnetic field of $H=0.2\;\text{T}$. Bare Nb and PdAu-capped Nb/PdAu have similar critical current densities, whereas Au-capped Nb/Au and Nb/Au-epi both have lower values. This is consistent with our magneto-optical imaging and magnetization measurements. A higher critical current means that the flux pinning is stronger, so the flux penetration depth is shorter; this is what we observe in Fig.~\ref{fig:MO_ZFC_highT} at elevated temperatures, where we see that dendritic flux patterns no longer appear and uniform distributions can be studied. 

\begin{figure}[!tbh]
\includegraphics[width=0.95\linewidth]{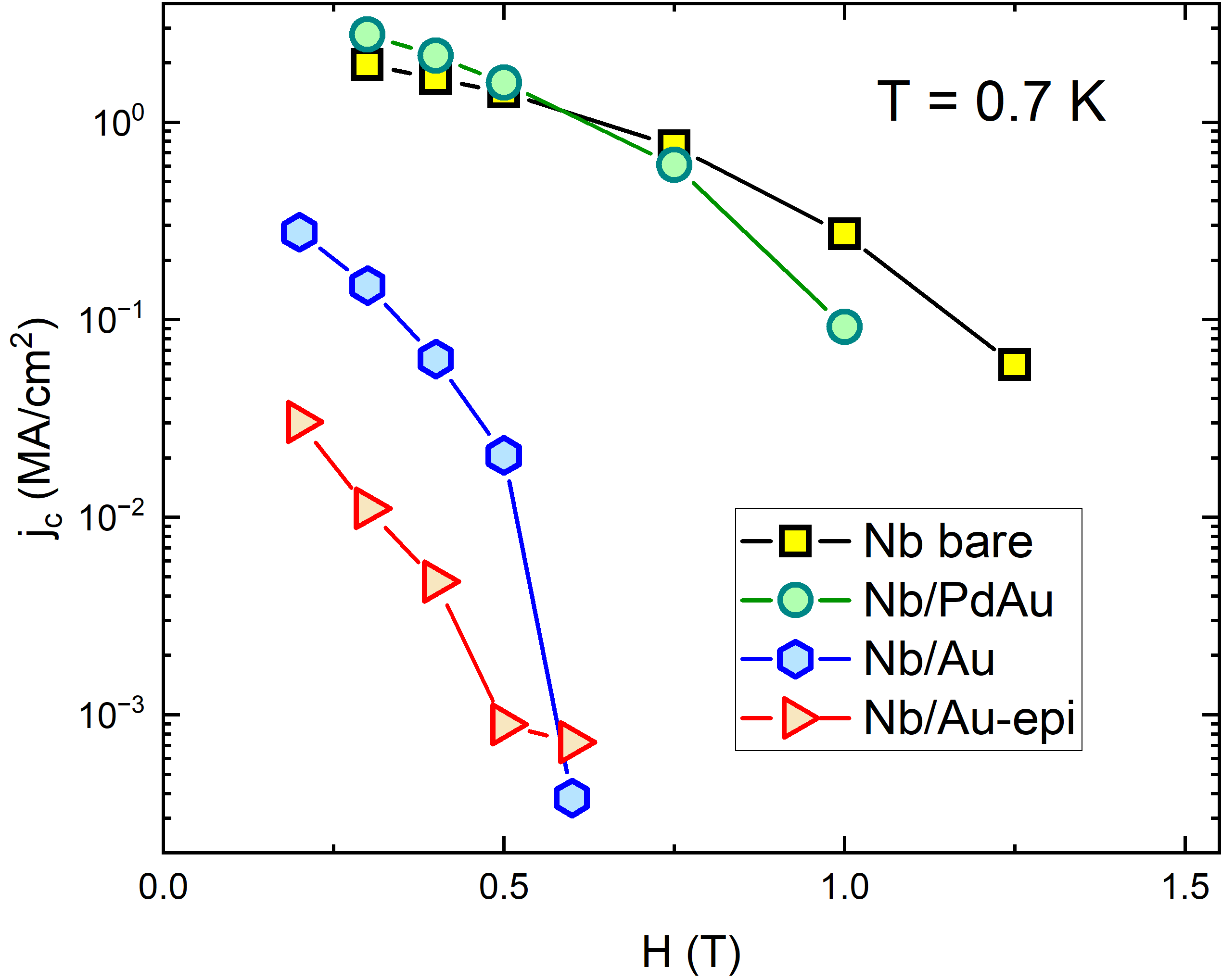} 
\caption{\textbf{Critical current density, $j_c(H)$.} Magnetic field dependence of $j_c(H)$ evaluated at $T=0.7$~K plotted on a logarithmic vertical scale.}
\label{fig:jc_H} 
\end{figure}

Figure~\ref{fig:jc_H} shows the dependence of critical current densities $j_{c}$ on the magnetic field. At $T=0.7\;\text{K}$, we are examining the values of $j_{c}$ for different magnetic fields, and we observe lower $j_{c}$ and lower $H_{c2}$ for Au capped films, which is consistent with Fig.~\ref{fig:Hc2_Tc}, where we have shown lower $H_{c2}(0)$ for Au capped films and higher values for bare and PdAu capped films. The difference in the strength of flux pinning is responsible for this distinct difference in behavior. The lower $H_{c2}(0)$ corresponds to a smaller scattering rate $\Gamma$, hence weaker vortex pinning.

\section{Discussion}

Surface encapsulation of thin Nb films is expected to suppress the formation of \ce{NbO_x}, at the surface upon exposure to air. The detrimental role of the oxides is well researched and documented \citep{Leon2021,McDermott2009,Oliver2013,Wang2015,Dial2016,Read2023,goronzy2025}. However, direct measurements, such as tunneling spectroscopy, conclude that the effect is confined to the surface layer and that the bulk remains largely unaffected  \citep{berti2023, rice2025, makita2024scanning}.

In this work, we present a set of independent measurements that show a significant change in all bulk superconducting and normal-state properties of metal-encapsulated niobium films. The normal-state property, the residual resistivity ratio, $RRR$, shows a dominant contribution of scattering due to defects and impurities, as opposed to electron-electron and electron-phonon scattering channels. The highest $RRR \approx 34$ is achieved in the epitaxial Au-capped film, followed by $RRR \approx 24$ in the sputtered Au-capped film.
The rest are significantly lower, ranging from 5 to 8, with the lowest in the bare Nb film, $RRR \approx 5$. These results indicate that Au-capped films are the cleanest.

This conclusion is further supported by comparing resistivity at T$_{c}$ with the transition temperature. The trend shown in Fig.\ref{fig:resistivity}(c) reveals monotonic behavior, with bare Nb exhibiting the highest resistivity and the lowest $T_c$, while Au-capped films are closer to the clean limit. The $T_{c}$ suppression due to nonmagnetic disorder arises from the anisotropy of the order parameter and the multiband nature of Nb \citep{Zarea2023}. Of course, if magnetic defects are also present, they suppress $T_{c}$ even more strongly.

Similarly, the upper critical field, $H_{c2}$, follows a monotonic trend, with the highest value in bare Nb and the lowest in epitaxially grown Nb/Au-epi, which also approaches the clean limit preceded by a sputtered Nb/Au film. The zero-temperature London penetration depth, $\lambda(0)$, exhibits a linear dependence on $\Gamma$. The Campbell length in Au-capped films exhibits higher values than those in bare Nb and PdAu-capped Nb, which in turn is reflected in a higher critical current density, $j_{c}$, in these films. 

Other affected properties include the heat conduction to the substrate. The observed thermo-magnetic instabilities manifest themselves in the form of dendritic magnetic flux avalanches, as shown by magnetization (Fig.~\ref{fig:magnetization_PdAu}) and direct magneto-optical visualization (Fig.\ref{fig:MO_ZFC_lowT} and Fig.\ref{fig:MO_FC}). There is a significant difference between the studied films. The avalanches are observed in sputtered bare-Nb and Nb/PdAu-capped films. Gold-capped films do not show these features, which is consistent with the smaller critical current, possibly due to the inhibition of oxygen and nitrogen diffusion from the surface. 

At elevated temperatures, the instabilities subside, as shown in Fig.\ref{fig:MO_ZFC_highT}. However, transmon qubits operate at ultra-low temperatures, and our observation of thermomagnetic instabilities reveals insufficient thermal channeling to the substrate, which could be an issue at the GHz microwave frequencies of their operation. Additionally, this may also indicate that issues exist at the atomic scale at the interface with the substrate. 

All our results may be understood in terms of different amounts and types of disorder in the bulk of the films. To be effective in the suppression of $T_c$ and in increasing $j_c$, disorder must vary from atomic defect size to coherence length scale, up to 20 nm in the studied films. To investigate possible sources, we performed time-of-flight secondary ion mass spectrometry (TOF-SIMS) measurements on all sputtered films. With this technique, the depth profiles of impurities throughout the film thickness were examined, as presented in detail in previous published works \citep{Murthy2022,oh2024}. In our particular study, the bare Nb film has the highest oxygen concentration, exhibiting almost ten times higher counts of oxygen than that of the Au-capped sputtered film, which is directly correlated with the highest disorder scattering and the lowest observed $T_{c}$. Not all capping layers behave equally. Although PdAu capping suppresses oxygen diffusion, the PdAu-capped film shows the highest NbN concentration, with 2 times higher counts compared to the Au-capped sputtered film, which could also act as a source of disorder, introducing pinning and suppressing $T_{c}$. However, the reason for the elevated NbN content is likely extrinsic—this film was deposited in a chamber also used for NbN. Pre-sputter cleaning of the target is not sufficient to obtain pure Nb. Both Au-capped sputtered Nb films have the lowest oxygen concentration and a low amount of NbN. This likely explains why Au capped films show low pinning and higher $T_{c}$, with surface oxide suppression being the primary factor.

In conclusion, we demonstrated that the actively used surface metal-encapsulation affects not only the surface but also all bulk properties of thin Nb films. A better microscopic understanding of the mechanisms behind these effects may lead to significant progress in solving the quantum decoherence problem in 2D superconducting qubits.

\section{Methods}

\subsection{Niobium thin films} \label{section:samples}

Niobium films for this study were fabricated on sapphire substrates using either sputtering or molecular beam epitaxy. For each type, we studied both control (reference) uncapped films and Au-coated films. The Nb/PdAu-capped film was fabricated by sputtering. Details of sample preparation are described elsewhere \citep{Bal2024,Murthy2025}.  It is noted that the capping covers only the top surfaces, and the sidewalls retain their nominal oxide layers. 

In this study, six niobium film samples were used. 

\begin{itemize}
\item \ul{Bare Nb}: un-capped sputtered Nb film (163\,nm Nb)
\item \ul{Nb/PdAu}: sputtered Nb film, followed by the sputtering of a PdAu layer, all under high vacuum ($\sim 10^{-8}$\,torr) (155\,nm Nb + 6\,nm PdAu)
\item \ul{Nb/Au}: sputtered Nb film was removed from the sputtering chamber and transferred into an e-beam chamber. At this stage, Ar$^+$ ions were
used to remove the surface layer, presumably minimizing the oxides, followed by the electron-beam deposition of an Au overlayer. (150\,nm Nb + 9\,nm Au)
\item \ul{Nb/Au-2}: similar to the one above, but with different deposition parameters, which included annealing the film at \SI{1100}{\celsius} for \SI{2}{\hour}, resulting in a significant difference in properties.  (144\,nm Nb + 10\,nm Au)
\item \ul{Bare Nb-epi}: molecular beam epitaxy - deposited blank
niobium film. (120\,nm Nb)
\item \ul{Nb/Au-epi}: MBE-deposited Nb film followed by in-situ Au encapsulation,
without breaking the vacuum. (110\,nm Nb + 5\,nm Au)
\end{itemize}

It should be noted that sputtered PdAu-capped and Au-capped films were fabricated in different sputtering chambers, which always have unique residuals from other sputtered materials, so the impurities from the different systems are distinct. For example, chambers at NIST are also used for depositing NbN, and pre-sputtering cleaning of the target might be insufficient to obtain pure Nb. This is likely the reason for the higher content of NbN in some sputtered films, as reported in the discussion.
\subsection{Low-temperature magneto-optical imaging} \label{section:MO}

Optical studies were conducted using a low-temperature, linearly-polarized light setup that incorporated a closed-cycle optical cryo-station (Model-s50 from \textit{Montana Instruments}) and the \textit{Olympus BX3M} optical system with long focal length objectives located just above the cryostat. The sample was mounted on a gold-plated cold stage and could be directly observed through optical windows. This fully closed-cycle $^{4}$He system enables controlled measurements across a wide temperature range, from room temperature to 3.8 K.

The two-dimensional distribution of magnetic induction on the sample surface was mapped in real-time using magneto-optical (MO) imaging, which leverages the Faraday effect in transparent bismuth-doped iron-garnet ferrimagnetic indicators placed directly on top of the flat sample surface. Linearly polarized light acquires a double Faraday rotation
when propagating through the indicator and reflecting back from the mirror layer at the bottom of the indicator. The magnetic field on the sample surface polarizes the in-plane magnetic moments of the indicator, and this spatially resolved polarization is visualized as a 2D optical image. In all images, only the distribution of the magnetic induction
on the sample surface is visualized since the light never reaches the sample surface itself. Brighter areas correspond to regions of higher magnetic induction, while dark areas indicate complete magnetic screening (B = 0). Further details on this technique can be found in earlier studies \citep{Prozorov2006a, Datta2024, Joshi2023, Jooss2002}.

\subsection{Tunnel-diode resonator: magnetic penetration depth} \label{section:TDR}

Magnetic penetration depth was measured using a highly sensitive self-oscillating
tunnel diode resonator (TDR) \citep{VanDegrift1975RSI,Prozorov2000,Prozorov2000a,Prozorov2006,Prozorov2011,Prozorov2021}. The sample is placed inside a single-layer inductor coil that generates a small AC magnetic field $H_{ac}<2\;\mu\text{T}$ at its resonant frequency of approximately $f_{0} \approx 14$~MHz. A tunnel diode is connected in series to the $LC-$circuit. When the diode is biased to the region of its negative differential resistance, the tank circuit starts resonating when the impedances match, and the losses in the circuit are compensated. When the system is well-stabilized and isolated, it achieves a resolution better than 1 part per billion, capable of detecting frequency shifts as small as 0.01 Hz relative to the 14 MHz base frequency. The measurement is similar to the microwave cavity perturbation technique; however, TDR is always locked to its resonance frequency and resonates at the radio frequency.

Changes in the sample magnetic response cause a change in the total inductance, resulting in a frequency shift, $\Delta f/f_{0}$, which is proportional to the magnetic susceptibility, $\chi\left(T\right)$, with a sample-specific calibration constant expressed as $\Delta f\left(T\right)/f_{0}=G\chi\left(T\right)$. Here $f_{0}$ is the empty resonator frequency. This susceptibility can be further converted into the magnetic penetration depth using the relation $\left(1-N\right)\chi=\lambda/R\tanh\left(R/\lambda\right)-1$,
where $N$ is the generalized demagnetizing factor \citep{Prozorov2018},
and $R$ is the effective sample dimension \citep{Prozorov2021}.
At low temperatures (below $0.8T_{c}$), where $\tanh\approx1$, a
simplified relation $\chi\left(T\right)\sim\lambda\left(T\right)$
can be applied. More details of the TDR technique and its application to measure
magnetic penetration depth are given elsewhere \citep{Prozorov2000a,Prozorov2006,Prozorov2011,Carrington2011,Prozorov2021,Giannetta2022}. 

Without an applied external DC magnetic field, the sample is in the Meissner state with no Abrikosov vortices. In this case, the measured magnetic penetration depth is the London penetration depth, $\lambda_{m}=\lambda$. Knowing $\lambda$ allows for the evaluation of the superfluid density,
$\rho_{s}=\left(\lambda\left(0\right)/\lambda\left(T\right)\right)^{2}$, which can be analyzed to determine the superconducting gap structure and quasiparticle spectrum \citep{Joshi2023}. When an external magnetic field is applied, the total magnetic penetration depth contains both the London and Campbell penetration depths, and $\lambda_{m}^{2}=\lambda^{2}+\lambda_{C}^{2}$
\citep{Koshelev1991,Brandt1991,Brandt1995,Brandt1998a}. The Campbell
penetration depth provides important information about the structure of the pinning potential and the unrelaxed critical current density \citep{Campbell1969,Campbell1971,Kim2021,Ghimire2025}.

\subsection{Electrical resistivity}

Four-probe electrical resistivity measurements were performed in a \emph{Quantum Design} PPMS\footnote{\label{fn:PPMS}Certain commercial equipment, instruments, software, or materials, commercial or non-commercial, are identified in this paper in order to specify the experimental procedure adequately. Such identification does not imply recommendation or endorsement of any product or service by NIST, nor does it imply that the materials or equipment identified are necessarily the best available for the purpose.} Contacts were made by gluing 25 $\mu$m silver wires using DuPont~4929N conducting silver paste. This technique yields contacts with contact resistance in the 10 to 100~$\Omega$ range. Transport measurements of the upper critical field, $H_{c2}$,
were performed with a magnetic field oriented perpendicular to the film plane to avoid the third critical field, $H_{c3}$, which does not exist in this orientation but is maximum when the magnetic field is parallel to the film surface \citep{Abrikosov2017,Kogan2002}. (The demagnetizing corrections are irrelevant near $H_{c2}$.) For measurements in a parallel configuration, the sample was glued to the side of a cube, similar to Ref.~\citep{YongLiu}. This procedure provided alignment with an accuracy of about 2 degrees.

\subsection{Magnetization}

The magnetic moment was measured using a commercial \textit{Quantum Design} (QD) vibrating sample magnetometer (VSM) in a $9\:\text{T}$ \textit{Physical Property Measurement System} (PPMS)\ref{fn:PPMS}. The VSM operates at a low frequency of 40 Hz in a static magnetic field, making it a very good approximation of true DC magnetization.
To avoid demagnetizing effects, the magnetic field was oriented parallel to the film surface. The demagnetizing effects are evident at small magnetic fields when the same sample is measured in two orientations, and the transition curve, $M\left(T\right)$,
in the perpendicular orientation is broader than that in the parallel orientation, indicating the nucleation of vortices.

\section*{Data availability}

Data are available upon reasonable request.

\section*{Author contributions}

R.P. coordinated the project. 
The FNAL team (M.B, Z.S., S.G., F.C., A.M., A.R., A.G) fabricated, coated, and characterized sputtered films.
The NWU team (D.G., D.P.G, M.C.H., M.J.B.) fabricated, coated, and characterized epitaxial films.
The NIST team (D.O. and P.H) fabricated sputtered films.
A.D. and K.R.J. performed magneto-optical imaging. 
A.D., K.R.J. and S.G. performed London penetration depth measurements. 
B.S.M. and M.A.T. performed transport measurements. 
R.P. performed vibrating sample magnetometry measurements. 
All authors from FNAL, NIST, NWU, and Ames contributed to the preparation, characterization, and compositional analysis of the studied films. 
All authors contributed to the interpretation of the results and wrote the manuscript.

\section*{Competing interests}

The authors declare no competing interests.

\section*{Correspondence}

Correspondence and requests for materials should be addressed to Ruslan Prozorov at prozorov@ameslab.gov

\begin{acknowledgments}
We thank James Sauls, Mehdi Zarea, Maria Iavarone, and John Zasadzinski for numerous fruitful discussions.
This work was supported primarily by the U.S. Department of Energy, Office of Science, National Quantum Information Science Research Centers, Superconducting Quantum Materials and Systems Center (SQMS), under Contract No. 89243024CSC000002. Ames Laboratory is supported by the U.S. Department of Energy (DOE), Office of Science, Basic Energy Sciences (BES), Materials Science \& Engineering Division (MSED) and is operated by Iowa State University for the U.S. DOE under contract DE-AC02-07CH11358. M.A.T. and K.R.J. were supported by DOE, BES, MSED at Ames National Laboratory. Fermilab is operated by Fermi Forward Discovery Group, LLC under Contract No. 89243024CSC000002 with the U.S. Department of Energy, Office of Science, Office of High Energy Physics.  

\end{acknowledgments}


%

\end{document}